\documentclass[preprint,tightenlines,showpacs,floatfix,nofootinbib,
superscriptaddress,fleqn]{revtex4}
\usepackage{graphicx}
\usepackage{graphicx}
\usepackage{bm}
\usepackage{dcolumn}
\usepackage{amsmath}
\usepackage{amsfonts}
\usepackage{amssymb}

\begin{document}

\bibliographystyle{prsty}


\preprint{PNU-NTG-08/2005}
\preprint{RUB-TPII-02/2005}
\title{Axial-vector form factors of the nucleon within the
chiral quark-soliton model and their strange components}

\author{Antonio Silva}
\email{ajose@teor.fis.uc.pt}
\affiliation{Centro de Fisica Computacional, Departamento de Fisica da 
Universidade de Coimbra, P-3004-516 Coimbra, Portugal}

\author{Hyun-Chul Kim}
\email{hchkim@pusan.ac.kr}
\affiliation{Department of Physics and Nuclear Physics \& Radiation
Technology Institute (NuRI), Pusan National University,
609-735 Busan, Republic of Korea}

\author{Diana Urbano}
\email{urbano@fe.up.pt}
\affiliation{Centro de Fisica Computacional, Departamento de Fisica da 
Universidade de Coimbra, P-3004-516 Coimbra, Portugal}
\affiliation{Faculdade de Engenharia da Universidade do Porto,
R. Dr. Roberto Frias s/n, P-4200-465 Porto, Portugal} 

\author{Klaus Goeke}
\email{Klaus.Goeke@tp2.rub.de}

\affiliation{Institut f\"ur Theoretische  Physik II,
Ruhr-Universit\" at Bochum, D--44780 Bochum, Germany}
\date{September 2005}

\begin{abstract}
We investigate three different axial-vector form factors of the
nucleon, $G_A^{0}$, $G_A^3$, $G_A^8$, within the framework of the
SU(3) chiral quark-soliton model, emphasizing their strangeness
content. We take into account the rotational $1/N_c$ and linear
strange quark ($m_s$) contributions using the symmetry-conserving
SU(3) quantization and assuming isospin symmetry. The strange
axial-vector form factor is also obtained and they all are
discussed in the context of the parity-violating scattering of
polarized electrons off the nucleon and its relevance to the
strange vector form factors.
\end{abstract}

\pacs{12.40.-y, 14.20.Dh}
\maketitle

\section{Introduction}
Axial-vector properties of the nucleon provide deep insight in
understanding its internal structure.  For instance, the measured
first moment $I_p$ of the proton spin structure
function~\cite{EMC,SMC1,SMC2,E142,E143,Abe:1997dp,Airapetian:1998wi,
Anthony:1999rm,Anthony:2000fn} is directly related to the quark
content of the nucleon at a fixed $Q^2$:
\begin{equation}
\Gamma_1^p (Q^2) = \int_0^1 g_1^p(x,Q^2) dx =
\frac{C_{ns}(Q^2)}{12}a_3 +\frac{C_{ns}(Q^2)}{36} a_8 +
\frac{C_s(Q^2)}{9} a_0,
\end{equation}
where $g_1^p(x)$ denotes the spin structure function of the
proton, and the $a_i$ are the singlet and non-singlet combinations
of the first moments of the parton distributions:
\begin{eqnarray}
&&a_0 =g_A^0= \Delta \Sigma = \Delta u + \Delta \bar{u} + \Delta d
+
\Delta \bar{d} + \Delta s + \Delta \bar{s},\notag \\
 &&a_3=g_A^3=\Delta u
+ \Delta \bar{u}
  -   \Delta d - \Delta \bar{d}, \notag \\
  &&a_8 =g_A^8= \Delta u + \Delta \bar{u}+
   \Delta d + \Delta \bar{d}- 2\Delta s - 2 \Delta \bar{s},
\label{eq:qspin}
\end{eqnarray}
and $C_s$ and $C_{ns}$ are the first moments of the non-singlet
coefficient functions which can be computed in perturbative QCD.
The first moments $a_i$ can be identified as the axial-vector
constants $g_A^i$ in the quark parton model.  In particular, the
$g_A^0=\Delta \Sigma$ is just interpreted in the quark parton
model as the fraction of the proton angular momentum carried by
the intrinsic spins of the quarks. The $g_A^3$ and $g_A^8$ are
related to semi-leptonic decays. Thus, the axial-vector constants
provide essential information on understanding the spin structure
of the nucleon.

Recently, parity-violating asymmetries of the polarized electrons
scattered off the nucleon have been intensively measured~\cite{
Mueller:1997mt,Spayde:2000qg,
SAMPLE00s,Aniol:2000at,Maas,happex,Aniol:2005zf,Aniol:2005zg,
Armstrong:2005hs}, since two neutral electroweak form factors
$G_E^A$ and $G_M^Z$ of the proton can be extracted from the
asymmetries. Once these form factors are known, one can combine
them with $G_E^\gamma$ and $G_M^\gamma$ and can  disentangle the
up, down, and strange quark contributions to the electromagnetic
form factors of the nucleon. However, it is also essential to know
information on the neutral weak axial-vector form factor, $G_A^Z$,
defined as
\begin{equation}
G_A^Z (Q^2) = -\tau_3 G_A (Q^2) + G_A^s(Q^2),
\end{equation}
where $G_A$ stands for the triplet axial-vector form factor of
which the value at $Q^2=0$ is just the axial-vector constant
$g_A^3$ determined from nucleon $\beta$ decay.  Note that $Q^2$ is
the positive four-momentum transfer, $Q^2=-q^2$. The $G_A^s$
denotes the strange axial-vector form factor and its value at
$Q^2=0$ is the strange quark content of the nucleon spin. These
axial-vector form factors have been experimentally investigated by
quasi-elastic neutrino-nucleon scattering \cite{Kitagaki:px} and
by pion electroproduction \cite{Liesenfeld:1999mv}. There a
dipole-type parametrization has been used to extract the
strange-vector form factors (see Ref.~\cite{Beise:2005wr} in
detail):
\begin{equation}
G_A (Q^2) = \frac{g_A}{(1+Q^2/M_A^2)^2},
\label{eq:dipol}
\end{equation}
where $M_A$ is the axial dipole mass.  Apparently, the
axial-vector properties of the nucleon shed light on the structure
of the nucleon in various aspects.

Though a correct theoretical description of the structure of the
nucleon should be based on QCD, which would automatically include
the excitation of $\rm q\bar{q}$ pairs in the nucleon, it is
rather difficult to perform in practice such a calculation. It is
even very hard to use lattice gauge techniques because they still
suffer from technical problems, in particular, in treating light
quarks. Thus, we need to use appropriate models which reflect
important characteristics of QCD such as chiral symmetry and its
spontaneous symmetry breaking and which treat the relevant degrees
of freedom in some good approximation.  In this sense, the chiral
quark-soliton model ($\chi$QSM) provides a proper theoretical
framework.  It is an effective quark theory of QCD in the limit of
$N_c \rightarrow \infty$ concentrating on the instanton-degrees of
freedom of the QCD vacuum and resulting in an effective action for
valence and sea quarks both moving in a static self-consistent
Goldstone background
field~\cite{Diakonov:1988ty,Christov:1996vm,Alkofer:1994ph}.  The
nucleon arises from this effective interaction as a chiral
soliton. It has successfully been applied to various properties of
the baryons~\cite{Christov:1996vm} and to forward and generalized
parton distributions~\cite{Diakonov:1996sr,Petrov:1998kf,
Goeke:2001tz} and has lead even to the prediction of the heavily
discussed pentaquark baryon $\Theta^+$~\cite{Diakonov:1997mm}.

The present authors have recently investigated the strange vector
form factors~\cite{Silva:2001st}, utilizing the
symmetry-conserving quantization scheme suggested by
Prasza{\l}owicz {\em et al.}~\cite{Praszalowicz:1998jm}.  We
predicted the SAMPLE, HAPPEX, A4, and G0
experiments~\cite{Silva:2001st,Silva:2002ej}. Results have shown a
fairly good agreement with experimental data of the A4, SAMPLE,
and HAPPEX.  However, since the experimental uncertainties are
rather large, it is hard to judge any theoretical calculation.
While the extracted strange vector form factors appear to have
large experimental uncertainties, the parity-violating asymmetries
are measured with relatively good precision.  Thus, the
parity-violating asymmetries may be more reliable to justify the
model descriptions.

In the present work, we want to study all three axial-vector form
factors within the framework of the $\chi$QSM, focusing on the
strangeness content of the nucleon.  In fact, axial-vector
properties of the nucleon have to some extent already been studied
in the SU(2) and SU(3) $\chi$QSM,
previously~\cite{Wakamatsu:ag,Wakamatsu:up,Christov:1993ny,
Alkofer:1993pv,Christov:1994ea,Watabe:1995ke,Blotz:1993dd,
Blotz:wi,Kim}.  However, these studies were not complete and some
of them did not use the symmetry conserving quantization
method~\cite{Praszalowicz:1998jm}, which guarantees that the
Gell-Mann-Nishijima relations are obeyed. Hence, we will calculate
all axial-vector form factors in this work, employing the
symmetry-conserving quantization.  We will follow also the same
set of parameters we used for deriving the strange vector form
factors so that the parity-violating asymmetries may be
studied~\cite{asym}.

This paper is organized as follows: In Section
II, we briefly show how to derive the axial-vector form factors in the
$\chi$QSM.  In Section III, we present the results and discuss them.
In Section IV, we summarize and draw the conclusion of the present
work.

\section{Axial-vector form factors in the $\chi$QSM}
In the present section we briefly mention the general expression
for the axial-vector form factors in the $\chi$QSM (see
also reviews~\cite{Diakonov:1988ty,Christov:1996vm}).

The axial-vector form factors are expressed in terms of the following
quark matrix elements:
\begin{equation}
\langle N (p',S_3')|A^{(\chi)}_{\mu }| N(p,S_3)\rangle \; =\;
\bar{u}_{N}(p',S_3') \left[\gamma _\mu\gamma_5
G_{A}^{(\chi)}(q^{2})+i\gamma_5 \frac{q^{\nu}}{2M_N}
G_{P}^{(\chi)}(q^{2})\right] u_{N}(p,S_3),
\label{Eq:ff1}
\end{equation}
where $G_A^{(\chi)} (q^2)$ and $G_P^{(\chi)}(q^2)$ are the
axial-vector and induced pseudoscalar form factors, respectively.
Note that we ignore the form factor of the second kind, {\em
  i.e.} the induced pseudotensor form factor, since it is irrelevant
to the present work. The $q^{2}$ is the square of the four
momentum transfer $q^{2}=-Q^{2}$ with $Q^{2}>0$. The $M_N$ and
$u_{N}(p, S_3)$ stand for the nucleon mass and its spinor with the
third component of the nucleon spin, respectively.  The
axial-vector quark current $A_{\mu}^{(\chi)}$ can be expressed in
Euclidean space as follows:
\begin{equation}
\label{Eq:scur}
A_{\mu}^{(\chi)} (x) \;=\; -i\psi^\dagger(x) \gamma _{\mu
}\gamma_5\frac{\lambda^\chi}{2} \psi (x),
\end{equation}
where $\lambda^{\chi}$ are the Gell-Mann matrices.

Since we are interested in the axial-vector form factors of the
nucleon in low-energy regions ($Q^2\le1\, {\rm GeV}^2$), we will
work in the Breit frame, where the time component of the
axial-vector current vanishes.  Thus, the axial-vector form
factors can be related only to the space component of the
axial-vector current. This is no real restriction since its time
component is suppressed in the large $N_c$ limit, compared to the
spatial one. Having carried out a tedious but straightforward
calculation following
Refs.\cite{Kim:1995mr,Christov:1996vm,Silva}, we derive the
expressions for the triplet ($\chi=3$) and octet ($\chi=8$)
axial-vector densities of the nucleon:

\begin{eqnarray}
G_{A}^{(\chi)}\ (q^{2}) &=&\frac{M}{E}\int d^{3}x\left[ j_{0}\left(  \left|
\bm{q}\right|  r\right)
\left\langle N(S_{3})\right|  A^{3\chi}(\bm{x})\left|
  N(S_{3})\right\rangle\right. \cr
& &-   \left.\sqrt{2\pi}j_{2}\left(  \left|  \bm{q}\right|
r\right) \left\langle N(S_{3})\right|  \left\{  \bm{Y}_{2}\left(
\hat{\bm{r}}\right) \otimes{\bm
A}_{1}^\chi(\bm{x})\right\}_{10}\left|
N(S_{3})\right\rangle \right] \nonumber \\
&= &\frac{M}{E}\int r^{2}dr\left[j_{0}\left(  \left|
\bm{Q}\right| r\right)
A_{0}^{\chi,B}(r)-\frac{1}{\sqrt{2}}j_{2}\left(  \left|
\bm{Q}\right| r\right)  A_{2}^{\chi,B}(r)\right], \label{eq:Axff}
\end{eqnarray}
where
\begin{eqnarray}
A_{0}^{\chi,B}(r)  & =&\int d\Omega\left\langle N(S_{3})\right|
A^{3\chi}({\bm x})\left|  N(S_{3})\right\rangle, \label{eq:A01} \\
A_{2}^{\chi,B}(r)  & =&\int d\Omega\left\langle N(S_{3})\right|
\left\{\sqrt{4\pi}{\bm Y}_{2}\left(  \hat{\bm r}\right)
  \otimes{\bm A}_{1}^\chi({\bm x})\right\}  _{10}\left|
  N(S_{3})\right\rangle\, .
\label{eq:A02}
\end{eqnarray}
Here $E$ is the on-shell energy of the nucleon, $E=\sqrt{M_N^2 +
{\bm q}^2/4}$. The $j_0(|{\bm Q}|r)$ and $j_2(|{\bm Q}|r)$ denote
the spherical Bessel functions. The $A^{3\chi}$ is the third
component of the axial-vector current in Eq.(\ref{Eq:scur}) and
the ${\bm Y}_2$ denotes the second-rank spherical tensors.

Having carried out the symmetry conserving collective quantization
with the linear $m_s$ corrections as well as rotational $1/N_c$
corrections taken into account, we end up with the following
expressions for Eq.(\ref{eq:A01}):
\begin{eqnarray}
&&\left\langle N(S_{3})\right| A^{3\chi=3,8}(\bm{z})\left|
  N(S_{3})\right\rangle   \cr
&= &
\frac{1}{3}\left\langle
  D_{\chi3}^{(8)}\right\rangle_N\mathcal{A}_0\left(  {\bm z}\right)
+\frac{1}{3\sqrt{3}I_{1}}\left(\left\langle
    J_{3}D_{\chi8}^{(8)}\right\rangle_N -2K_{1}m_{8}\left\langle
D_{83}^{(8)}D_{\chi8}^{(8)}\right\rangle_N \right)
\mathcal{B}_0\left(\bm{z}\right) \nonumber\\
&& +\frac{1}{3I_{2}}\left(  \left\langle d_{3ab}D_{\chi
      a}^{(8)}J_{b}\right\rangle_N -2K_{2}m_{8}\left\langle
    D_{8a}^{(8)}D_{\chi b}^{(8)} d_{ab3}\right\rangle_N \right)
\mathcal{C}_0\left(  \bm{z}\right)  -\frac
{i}{6I_{1}}\left\langle D_{\chi3}^{(8)}\right\rangle_N
\mathcal{D}_0\left(\bm{z}\right) \nonumber\\
&& +\frac{2}{3}\left(  m_{0}\left\langle
    D_{\chi3}^{(8)}\right\rangle_N +\frac{1}{\sqrt{3}}
m_{8}\left\langle D_{88}^{(8)}D_{\chi3}^{(8)}\right\rangle_N
\right)  \mathcal{H}_0\left(  \bm{z}\right) \nonumber\\
&& +\frac{2}{3\sqrt{3}}m_{8}\left\langle D_{83}^{(8)}D_{\chi8}^{(8)}
\right\rangle_N \mathcal{I}_0\left(  \bm{z}\right)
+\frac{2}{3} m_{8}\left\langle D_{8a}^{(8)}D_{\chi b}^{(8)}
d_{ab3}\right\rangle_N \mathcal{J}_0\left(  \bm{z}\right),\cr
&&\left\langle N(S_{3})\right|  A^{3\chi=0}(\bm{x})\left|
  N(S_{3})\right\rangle  \cr
&=&\frac{2}{3I_{1}}\left(  \left\langle J_{3}\right\rangle_N
-K_{1}m_{8}\left\langle D_{83}^{(8)}\right\rangle_N \right)
\mathcal{B}_0\left( \bm{z}\right)
+\frac{2}{3}m_{8}\left\langle D_{83}^{(8)}\right\rangle_N
\mathcal{I}_0\left(  \bm{z}\right) ,
\label{eq:expA0}
\end{eqnarray}
where $I_i$ and $K_i$ are the moments of inertia of the soliton.
The $m_0$ and $m_8$ denote the singlet and octet components of the
quark mass matrix: $m_0=(2\bar{m}+m_s)/3$ and
$m_8=(\bar{m}-m_s)/\sqrt3$ with $\bar{m}=m_u=m_d$, respectively.
The $d_{ab3}$ is the totally symmetric tensor in SU(3). The
$D^{(8)}$ represent the SU(3) Wigner functions and
$\langle\rangle_N$ stand for their matrix elements sandwiched
between the collective nucleon wave functions.  These matrix
elements are finally expressed in terms of the SU(3)
Clebsch-Gordan coefficients.  Note that here we tacitly have
considered the wave function corrections arising from
SU(3)-symmetry breaking, which mixes the pure nucleon state with
states in higher representations.  The nucleon state $J_3$ is the
third component of the spin operator.  The explicit expressions
for the quark densities, $\mathcal{A}_0$, $\mathcal{B}_0$,
$\mathcal{C}_0$, $\mathcal{D}_0$, $\mathcal{H}_0$,
$\mathcal{I}_0$, and $\mathcal{J}_0$ can be found in Appendix.

Note that the matrix element $\left\langle N(S_{3})\right| \{
\bm{Y}_{2}\left( \hat{\bm{r}}\right)
\otimes\bm{A}_{1}^{\chi=0}(\bm{x}) \} _{10}\left|
N(S_{3})\right\rangle $ in Eq.(\ref{eq:A02}) can be put into the
same form as Eq.(\ref{eq:expA0}), except for replacing the
operators in Eq.(\ref{eq:expA0}) by others which are given in
Table~\ref{tab:A0toA2}. Actually, in flavor SU(2), the above SU(3)
expressions are reduced to the isoscalar and isovector ones as
follows:
\begin{eqnarray}
A_{0}^{\chi=0}(r)  & = & A_{0}^{T=0} = \frac{1}{3I_{1}}\left\langle
  J_{3}\right\rangle_N  \mathcal{B}_{0}\left(  r\right), \\
A_{0}^{\chi=3}(r)  & = & A_{0}^{T=3} =  -\frac{1}{\sqrt{3}}\left\langle
  D_{33}^{(3)}\right\rangle_N
\mathcal{A}_{0}\left(  r\right)  -\frac{1}{3\sqrt{2}I_{1}}\left\langle
D_{33}^{(3)}\right\rangle_N \mathcal{D}_{0}\left(  r\right),
\end{eqnarray}
where the operators required for the function $A_2(r)$ in SU(2) can be
read out from Table~\ref{tab:A0toA2}.
\begin{table}[ht]
\begin{center}\begin{tabular}{rcl}               \hline\hline
Densities                            &      &   Operators  \\ \hline
$\mathcal{B}_0$, $\mathcal{I}_0$   &      &  ${\bm \sigma}$                                      \\
$\mathcal{B}_2$, $\mathcal{I}_2$   &      &
$\left\{\sqrt{4\pi}Y_2\otimes\sigma_1\right\}_1$   \\ \hline
$\mathcal{A}_0$, $\mathcal{C}_0$,  $\mathcal{H}_0$,  $\mathcal{J}_0$
                                   &      &
                                   ${\bm\sigma}\cdot{\bm\tau}$
                                   \\
$\mathcal{A}_2$, $\mathcal{C}_2$,  $\mathcal{H}_2$,  $\mathcal{J}_2$
                                   &      &
                                   $-\sqrt{3}\left\{
\left\{\sqrt{4\pi}Y_2\otimes\sigma_1\right\}_1\otimes\tau_1\right\}_0$
     \\ \hline
$\mathcal{D}_0$                    &      &
${\bm\sigma}\times {\bm\tau}$      \\
$\mathcal{D}_2$                    &      &
$-i\sqrt{2}\left\{\left\{\sqrt{4\pi}Y_2\otimes\sigma_1\right\}_1
\otimes\tau_1\right\}_1$            \\ \hline\hline
\end{tabular}\end{center}
\caption{\label{tab:A0toA2}%
This table shows how to obtain $A_2^{\chi,B}(r)$ by replacing
operators.  The factor $-\sqrt{3}$ comes from
${\bm\sigma}\cdot{\bm\tau}=-\sqrt{3}\left\{\sigma_1\otimes\tau_1\right\}_0$
and $-i\sqrt{2}$ from ${\bm\sigma}\times{\bm\tau}
=-i\sqrt{2}\left\{\sigma_1\otimes\tau_1\right\}_1$.}
\end{table}

Now, we turn to the flavor decomposition of the axial-vector form
factors.  The flavor axial-vector form factors can be expressed in
terms of the singlet and non-singlet ones as follows:
\begin{eqnarray}
G_A^u &=& \frac13 G_A^0 + \frac12 G_A^3 + \frac1{2\sqrt{3}} G_A^8,\cr
G_A^d &=& \frac13 G_A^0 - \frac12 G_A^3 + \frac1{2\sqrt{3}} G_A^8,\cr
G_A^s &=& \frac13 G_A^0 - \frac1{\sqrt{3}} G_A^8.
\label{eq:flavorff}
\end{eqnarray}
The fractions of the nucleon angular momentum carried by the
intrinsic spin of the quarks with given flavors are just
identified as the flavor components of the axial-vector form
factors at $Q^2=0$:
\begin{equation}
  \label{eq:fpart}
\Delta u+\Delta \bar{u} = G_A^u (Q^2=0), \;\;\; \Delta d+\Delta
\bar{d} = G_A^d (Q^2=0), \;\;\; \Delta s+\Delta \bar{s} = G_A^s
(Q^2=0).
\end{equation}

\section{Results and discussion}
We present now the results obtained from the $\chi$QSM.  A
detailed description of the numerical techniques to solve the
model can be found, for example, in
Refs.~\cite{Kim:1995mr,Christov:1996vm}. The parameters of the
model comprise the constituent quark mass $M$, the current quark
mass $m_{\rm u}=m_{\rm d}$, the cut-off $\Lambda$ of the
proper-time regularization, and the strange quark mass $m_{\rm
s}$. These parameters are not free but are adjusted to independent
observables in a very clear way.  In fact, this was done many
years ago: The $\Lambda$ and the $m_{\rm u}$ are fixed to a given
$M$ in the mesonic sector.  The physical pion mass $m_\pi = 139$
MeV and the pion decay constant $f_\pi = 93$ MeV are reproduced by
these parameters.  The strange quark mass is chosen to be $m_{\rm
s} = 180$ MeV throughout the present work.  The remaining
parameter $M$ is varied from $400$ MeV to $450$ MeV. However, the
value of $420\mbox{ MeV}$, which for many years is known to
produce  the best fit to many baryonic
observables~\cite{Christov:1996vm}, is selected for the present
result.

\begin{figure}[ht]
\centering
\includegraphics[height=7cm]{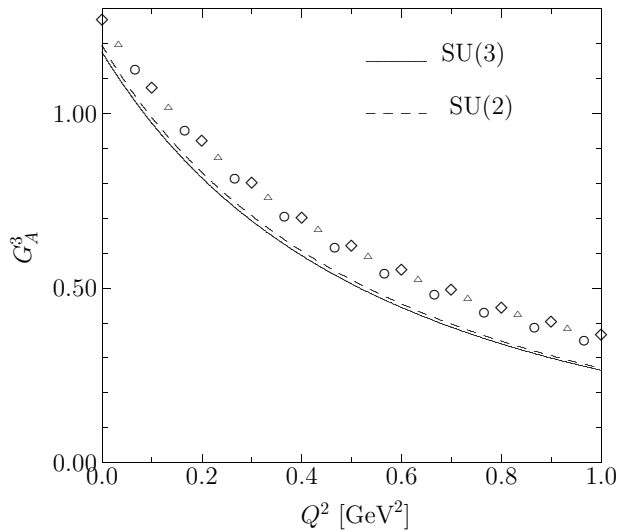}
\caption{The triplet axial-vector form factor, $G_A^3$ as a
function of $Q^2$.   The solid curve depicts the result of the
$\chi$QSM in the SU(3) version, while the dashed one draws $G_A$
from the SU(2) calculation.  The constituent quark mass is $M=420$
MeV. The open circles, squares, and triangles represent the
results with the dipole-type parametrization of the $G_A^3$ with
given empirical values of the axial-vector mass $M_A$.}
\label{fig:GA3}
\end{figure}

Figure~\ref{fig:GA3} depicts the results of the triplet
axial-vector form factors $G_A=G_A^3$.  The solid curve draws $G_A
(Q^2)$ from the SU(3) model, whereas the dashed one does its SU(2)
result. The open circles, squares, and triangles represent the
results based on Eq.(\ref{eq:dipol}) with three different
well-known average axial-vector masses: $M_A=1.032\pm0.036$~GeV
from neutrino (antineutrino) scattering experiments on protons and
nuclei~\cite{Kitagaki:px}, $M_A=1.069\pm0.016$~GeV from pion
electroproduction, and $M_A=1.077\pm0.039$ from charged pion
electroproduction \cite{Liesenfeld:1999mv}, respectively.  In all
these experimental analyses, the value $g_A=1.2673\pm 0.0035$
known from $\beta$-decay \cite{Eidelman:2004wy} was used for the
value of $G_A$ at $Q^2=0$. The $G_A$ from the SU(3) $\chi$QSM does
almost coincide with that of the SU(2) one. The theoretical SU(3)
value is $g_A=1.176$. Thus the SU(3) result is different from the
experimental data by only $8\,\%$. Compared to the empirical
dipole type $G_A$, the results in Figure~\ref{fig:GA3} show a very
similar $Q^2$-dependence.

The axial-vector radii are defined as
\begin{equation}
\left<r^2_\chi \right>_A=-6  \left.\frac{1}{G_A^{(\chi)}(0)}
\frac{dG_A(Q^2)}{dQ^2}\right|_{Q^2=0},
\label{eq:axrad}
\end{equation}
from which we obtain the triplet axial-vector radius of the SU(3)
$\chi$QSM : $\left<r^2_3 \right>_A^{1/2}=0.732\,{\rm fm}$.  The
experimental data is given as $\left<r^2_3
\right>_A^{1/2}=0.635\pm 0.023\,{\rm
  fm}$\cite{Liesenfeld:1999mv}.  Thus, the triplet axial-vector radus
from the model is also comparable to the data with a deviation of
less than $15\%$. The axial-vector dipole mass can be obtained
from Eq.~(\ref{eq:axrad}) as follows:
\begin{equation}
M_A^2=12/\left<r^2_3 \right>_A .
\end{equation}
We obtain for the axial-vector dipole mass of SU(3) $\chi$QSM:
$M_A=0.934\,{\rm GeV}$. Compared to the empirical values mentioned
above, the axial-vector dipole masss from the SU(3) $\chi$QSM
calculation deviates from the exmpirical ones by about $\pm
(10\sim 15)\,\%$.

\begin{figure}[ht]
\centering
\includegraphics[height=7cm]{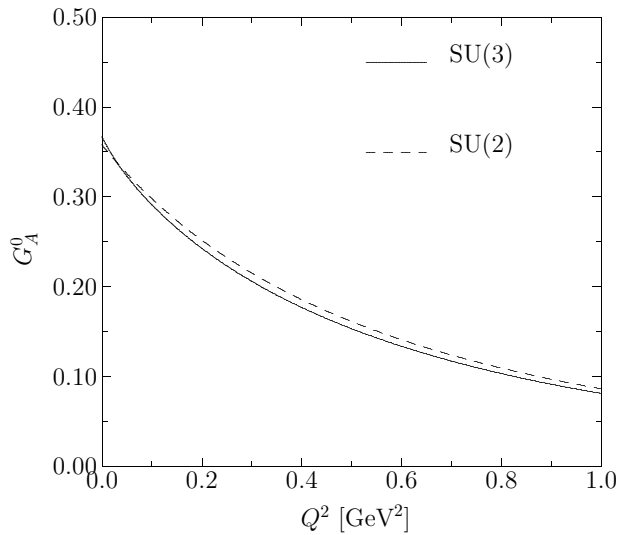}
\caption{The singlet axial-vector form factor, $G_A^0$ as a
function of $Q^2$.   The solid curve depicts the result of the
SU(3) $\chi$QSM model, while the dashed one draws $G_A$ from the
SU(2) calculation.  The constituent quark mass is $M=420$ MeV. }
\label{fig:GA0}
\end{figure}
\begin{figure}[ht]
\centering
\includegraphics[height=7cm]{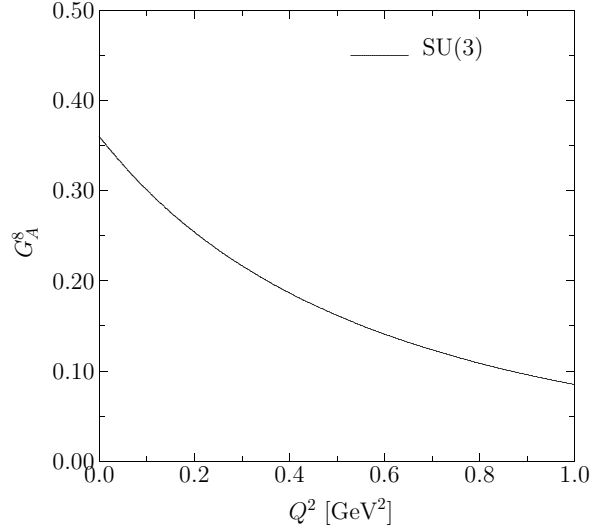}
\caption{The octet axial-vector form factor, $G_A^8$ as a function
of $Q^2$.   The solid curve depicts the result of the SU(3) model.
The constituent quark mass is $M=420$ MeV.}
\label{fig:GA8}
\end{figure}

Figure~\ref{fig:GA0} draws the results of the singlet axial-vector
form factors.  The general tendency is similar to the case of
$G_A^3$.  The SU(2) result is almost the same as that from the SU(3)
$\chi$QSM as in the case of $G_A^3$.  Figure~\ref{fig:GA8} represents
the results of the octet axial-vector form factors.  Note that in
SU(2) $G_A^8$ is not defined.

The singlet axial-vector constant $g_A^0=G_A^0 (Q^2=0)$ provides
information on the nucleon angular momentum carried by the
intrinsic spin of the light quarks as defined in
Eq.(\ref{eq:qspin}). It can be determined from the proton and
neutron structure function. The $g_A^0$ can be extracted from
longitudinally polarized deep-inelastic lepton scattering
experiments and is known approximately \cite{Filippone:2001ux} to
be $0.2\sim 0.35$. However, note that $g_A^{0}$ includes in
general the gluonic content of the nucleon spin which is
experimentally not well known.  From the present results, we get
$g_A^0 = 0.37$.

As for the octet axial-vector constant, it is related to semileptonic
decays and known experimentally to be $g_A^8=0.338\pm0.015$
\cite{Goto:1999by}.  As shown in Fig.~\ref{fig:GA8}, we obtain
$g_A^8=0.36$ which is in a remarkable agreement with the experimental
data.

\begin{figure}[ht]
\centering
\includegraphics[height=7cm]{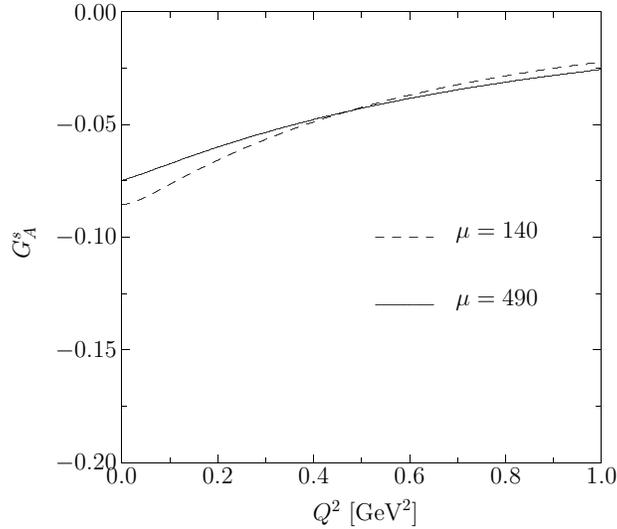}
\caption{The strange axial-vector form factor, $G_A^s$ as a function
of $Q^2$.   The solid curve and dashed curves depict the results
for the kaon ($\mu=490$ MeV) and pion ($\mu=140$ MeV) asymptotic
tails, respectively.  The constituent quark mass is $M=420$ MeV.}
\label{fig:GAs}
\end{figure}

Figure~\ref{fig:GAs} depicts the strange axial-vector form factor.
Actually two curves are given the difference of which is a measure
for the systematic error of $G_A^s (Q^2)$. The reasoning for that
has been discussed in Ref.~\cite{Silva:2001st,Silva:2002ej} and
can shortly be summarized as follows: The SU(3) soliton is
contructed by embedding the SU(2) one into the SU(3):
\begin{equation}
U_{\rm SU(3)}\; =\; \left( \begin{array}{cc}  U_{\rm SU(2)} & 0  \\
                                0 & 1
\end{array}\right) ,
\label{eq:imbed}
\end{equation}
with the SU(2) hedgehog given by
\begin{equation}
U_{\rm SU(2)}\; =\; \exp {[i\gamma_5 {\bf n}\cdot {\bm \tau
}P(r)]}\, . \label{eq:profile}
\end{equation}
The profile function $P(r)$ is found by minimizing the classical
energy of the soliton in a self-consistent way.  While the SU(2)
soliton incorporates the asymptotic pion behavior $\exp(-m_\pi
r)/r$ naturally, the SU(3) hedgehog by the embedding the SU(2)
soliton renders all other pseudo-Goldstone bosons have the same
asymptotic behavior by construction.  Therefore, this treatment is
phenomenologically unsatisfactory, in particular, when we deal
with quantities related to solely strange quarks. Thus, we will
consider here the kaonic asymptotic tail for the profile function
and look at the interval spanned by results of the more consistent
pion asymptotics and the phenomenologically driven kaon
asymptotics of $P(r)$ as giving an idea of the systematic model
uncertainties steming from the lack of exact treatment of the
SU(3) meson asymptotics. It is interesting to see that for $G_A^s
(Q^2)$ the result with the kaon tail does not differ much from
that with the pion one, which was not the case for the strange
vector form factors $G_M^s (Q^2)$ and $G_E^s (Q^2)$
~\cite{Silva:2001st,Silva:2002ej}. It implies that the strange
axial-vector form factor is insensitive to the asymptotic part of
the axial-vector densities.

\begin{table}[ht]
\begin{center}\begin{tabular}{lrcr}      \hline\hline
                              &   SU(2)\hspace{1cm}     &
                              SU(3)\hspace{1cm}
                              &  \cite{Filippone:2001ux}       \\
                               \hline
$\Delta u+\Delta \bar{u}$\hspace{1cm}     &  $0.777$\hspace{1cm} &
\ \ $0.814$\hspace{1cm}
                              &
                              $0.78\pm0.03$                  \\
$\left<r_u^2 \right>^{1/2}$\hspace{1cm}&  $0.665$\hspace{1cm}    &
\ \ $0.751$\hspace{1cm}
                              &
                              \\
$\Delta d+\Delta \bar{d}$\hspace{1cm}     &  $-0.419$\hspace{1cm}
& $-0.362$\hspace{1cm}
                              &
                              $-0.48\pm0.03$                 \\
$\left<r_d^2 \right>^{1/2}$\hspace{1cm}   &  $0.668$\hspace{1cm} &
\ \ $0.688$\hspace{1cm}
                              &
                              \\
$\Delta s+\Delta \bar{s}$\hspace{1cm}     &
--\hspace{1cm}\phantom{---} & $-0.086\ (-0.075)$\hspace{1cm}
                              &
                              $-0.14\pm0.03$                \\
$\left<r_s^2 \right>^{1/2}$\hspace{1cm}   &
--\hspace{1cm}\phantom{---} & $0.554\ (0.172)$\hspace{1cm}
                              &
                              \\
\hline\hline
\end{tabular}\end{center}
\caption{\label{tab:duddds}%
Results for the fractions of the nucleon carried by quarks and
flavor axial-vector radii.  Model parameters are $M=420$~MeV and
$m_s=180$~MeV. The calculations are performed with the profile
function having a pion-tail. For the purely strange quantities the
results with the kaon tail are given in brackets for comparison.}
\end{table}

In Table~\ref{tab:duddds}, we list the results for each flavor
contribution to the nucleon spin and the flavor axial-vector
radii. Compared to the empirical results~\cite{Filippone:2001ux},
it is found that the results of this work for $\Delta q_i+\Delta
\bar {q}_i$ are slightly deviating from them by about $20\sim
25\,\%$ in general apart from $\Delta s+\Delta \bar{s}$.  However,
note that those data in Ref.~\cite{Filippone:2001ux} are extracted
assuming SU(3) symmetry. This assumption has been shown by some of
the present authors in Ref.~\cite{Kim} to cause error bars for
$\Delta s+\Delta \bar{s}$ very much larger than those given in
Ref.~\cite{Filippone:2001ux}.

\section{Summary and conclusion}
In the present work, we have investigated the singlet and
non-singlet axial-vector form factors within the framework of the
SU(3) chiral quark-soliton model, incorporating the
symmetry-conserving quantization.  The rotational $1/N_c$ and
strange quark mass $m_{\rm s}$ corrections were taken into
account.  We have used the same set of parameters for the present
investigation so that we may utilize the present results in a
future study on the parity-violating asymmetries of the polarized
eletron scatterings off the proton.  The results of the singlet
and triplet axial-vector form factors turned out to be almost the
same as that with those in SU(2).  In order to see the sensitivity
to the asymptotics of the soliton, we employed pion and kaon tails
for the strange axial-vector form factors, as was done in
Ref.~\cite{Silva:2001st,Silva:2002ej}.  While the strange vector
form factors are rather
sensitive~\cite{Silva:2001st,Silva:2002ej}, the axial-vector one
is not.

\begin{table}[ht]
\begin{center}\begin{tabular}{lrcrcrrr}      \hline\hline
                               & \hspace{0.3cm}  &   SU(2)  &
                                \hspace{0.3cm}   &
                                SU(3)         &
                                \hspace{0.3cm} & Exp. &  \\
                                \hline
$g_A^0$                         & & $0.358$   & & $0.367$        &
                                 &
                                &  \\
$\left<r_0^2 \right>^{1/2}$     & & $0.662$   & & $0.844$        &
                                 &
                                &  \\
$M_A^0$                         & & $1.033$   & & $0.810$        &
                                 &
                                &  \\  \hline
$g_A^3$                         & & $1.196$   & & $1.176$        &
                                 &  $1.267\pm0.004$
                                &  \cite{Eidelman:2004wy} \\
$\left<r_3^2 \right>^{1/2}$     & & $0.666$   & & $0.732$        &
                                 &  $0.635\pm0.023$
                                &  \cite{Liesenfeld:1999mv}\\
$M_A$                         & & $1.026$   & & $0.934$        &
                                 &  $1.077\pm0.039$
                                &  \cite{Liesenfeld:1999mv}\\  \hline
$g_A^8$                         & & --        & & $0.360$        &
                                 &
                                &   \\
$\left<r_8^2 \right>^{1/2}$     & & --        & & $0.739$        &
                                 &
                                &  \\
$M_A^8$                         & & --        & & $0.926$        &
                                 &
                                &  \\  \hline
$g_A^s$                         &  & --       & & $-0.086\
(-0.075) $ &
                                  &  $-0.14\pm0.03$
                                & \cite{Filippone:2001ux} \\
$\left<r_s^2 \right>^{1/2}$     &  &--        & & $0.554\ (0.172)$
&
                                    &
                                &       \\
\hline\hline
\end{tabular}\end{center}
\caption{\label{tab:axconst}%
Axial-vector constants $g_A^\chi=G_A^{(\chi)}(Q^2=0)$, $g_A^s$,
axial-vector radii (in fm) (Eq.~\ref{eq:axrad}), and dipole
axial-vector masses (in GeV) from the dipole expression.  Model
parameters are $M=420$~MeV and $m_s=180$~MeV. The calculations are
performed with the profile function having a pion-tail. For the
purely strange quantities the results with the kaon tail are given
in brackets for comparison. }
\end{table}

The results of the axial-vector constants, corresponding radii,
and the dipole axial-vector masses are summarized in
Table~\ref{tab:axconst}.  The present results are in general good
agreement with the existing experimental data.

The present investigation posseses one noticeable virtue: Since we are
able to calculate the axial-vector form factors together with the
strange electromagnetic ones within the same scheme, we can directly
get access to the electroweak interactions of the proton.  In order to
calculate the parity-violating asymmetries of the polarized electrons
scattered off the proton, we must have all required form factors
derived consistently.  Actually, combining the present results of the
axial-vector form factors together with the strange vector form
factors, we can proceed to delve into the parity-violating asymmetries.
Since the parity-violating asymmetries have been and will be measured
in different kinematical regions with a good precision, they will
provide a criterion for theoretical works.

The parity-violating asymmetries of polarized electron-proton
scattering is presently under investigation.
\vspace{0.5cm}

\section*{Acknowledgments}
AS acknowledges partial financial support from the Portugese
Praxis XXI/BD/15681/98. The work has also been supported by
Korean-German grant of the Deutsche Forschungsgemeinschaft and
KOSEF (F01-2004-000-00102-0).  The work is partially supported by
the Transregio-Sonderforschungsbereich Bonn-Bochum-Giessen as well
as by the Verbundforschung and the International Office of the
Federal Ministry for Education and Research (BMBF).

\begin{appendix}
\section{Densities}
In this Appendix, we provide the densities for the axial-vector
form factors given in Eq.(\ref{eq:expA0}).  The sums run freely
over all single-quark levels including the valence one, except the sum
over $m_0$, which is restricted to negative-energy orbits:
\begin{eqnarray}
\frac{1}{N_{c}}\mathcal{A}_0\left(  {\bm z}\right)  &=&\left\langle
  {\rm val} \right. \left|  {\bm z}\right\rangle
  \bm{\sigma}\cdot\bm{\tau}\left\langle \bm{z}\right.  \left|
  {\rm val} \right\rangle +\sum_{n}
\mathcal{R}_{1}(\varepsilon_{n})\left\langle n\right.
\left|  \bm{z}\right\rangle
\bm{\sigma}\cdot\bm{\tau}\left\langle \bm{z}\right.  \left|n\right\rangle \\
\frac{1}{N_{c}}\mathcal{B}_0\left(  \bm{z}\right) & =&\sum_{m\neq0}\frac
{1}{\varepsilon_{{\rm val} }-\varepsilon_{m}}\left\langle {\rm val}
  \right.  \left| \bm{z}\right\rangle \bm{\sigma}\left\langle \bm{z}\right.  \left|
m\right\rangle \cdot\left\langle {\rm val} \right|  \bm{\tau}\left|
m\right\rangle \cr
 &&
-\frac{1}{2}\sum_{n,m}\mathcal{R}_{2}(\varepsilon_{n},\varepsilon
_{m})\left\langle n\right.  \left|  \bm{z}\right\rangle \bm{\sigma
}\left\langle \bm{z}\right.  \left|  m\right\rangle \cdot\left\langle
n\right|  \bm{\tau}\left|  m\right\rangle \\
\frac{1}{N_{c}}\mathcal{C}_0\left(  \bm{z}\right)   & =&\sum_{m^{0}}\frac
{1}{\varepsilon_{{\rm val} }-\varepsilon_{m^{0}}}\left\langle {\rm val} \right.  \left|
\bm{z}\right\rangle \bm{\sigma}\cdot\bm{\tau}\left\langle
\bm{z}\right.  \left|  m^{0}\right\rangle \left\langle m^{0}\right|
\left.  {\rm val} \right\rangle
\nonumber\\ &&
-\sum_{n,m^{0}}\mathcal{R}_{2}(\varepsilon_{n},\varepsilon_{m^{0}
})\left\langle n\right.  \left|  \bm{z}\right\rangle \bm{\sigma}
\cdot\bm{\tau}\left\langle \bm{z}\right.  \left|  m^{0}\right\rangle
\left\langle m^{0}\right|  \left.  n\right\rangle \\
\frac{1}{N_{c}}\mathcal{D}_0\left(  \bm{z}\right)   & =&\sum_{n}
\frac{\mathrm{sgn}(\varepsilon_{n})}{\varepsilon_{{\rm val} }-\varepsilon_{n}}\left\langle
n\right|  \left.  \bm{z}\right\rangle \bm{\sigma}\times\bm{\tau
}\left\langle \bm{z}\right|  \left.  {\rm val} \right\rangle \cdot\left\langle
{\rm val} \right|  \bm{\tau}\left|  n\right\rangle
 \nonumber\\ &&
+\frac{1}{2}\sum_{n,m}\,\mathcal{R}_{3}(\varepsilon_{n},\varepsilon
_{m})\left\langle m\right|  \left.  \bm{z}\right\rangle \bm{\sigma
}\times\bm{\tau}\left\langle \bm{z}\right|  \left.  n\right\rangle
\cdot\left\langle n\right|  \bm{\tau}\left|  m\right\rangle \\
\frac{1}{N_{c}}\mathcal{H}_0\left(  \bm{z}\right)   & =&\sum_{n\neq0}\frac
{1}{\varepsilon_{{\rm val} }-\varepsilon_{n}}\left\langle {\rm val} \right.  \left|
\bm{z}\right\rangle \bm{\sigma}\cdot\bm{\tau}\left\langle
\bm{z}\right.  \left|  n\right\rangle \left\langle n\right|  \gamma
^{0}\left|  {\rm val} \right\rangle
\nonumber\\ &&
+\frac{1}{2}\sum_{n,m}\mathcal{R}_{4}(\varepsilon_{n},\varepsilon
_{m})\left\langle m\right|  \left.  \bm{z}\right\rangle \bm{\sigma
}\cdot\bm{\tau}\left\langle \bm{z}\right|  \left.  n\right\rangle
\left\langle n\right|  \gamma^{0}\left|  m\right\rangle \\
\frac{1}{N_{c}}\mathcal{I}_0\left(  \bm{z}\right)   & =&\sum_{n\neq0}\frac
{1}{\varepsilon_{{\rm val} }-\varepsilon_{n}}\left\langle {\rm val} \right.  \left|
\bm{z}\right\rangle \bm{\sigma}\left\langle \bm{z}\right.  \left|
n\right\rangle \cdot\left\langle n\right|  \gamma^{0}\bm{\tau}\left|
{\rm val} \right\rangle
\nonumber\\ &&
+\frac{1}{2}\sum_{n,m}\mathcal{R}_{4}(\varepsilon_{n},\varepsilon
_{m})\left\langle m\right.  \left|  \bm{z}\right\rangle \bm{\sigma
}\left\langle \bm{z}\right.  \left|  n\right\rangle \cdot\left\langle
n\right|  \gamma^{0}\bm{\tau}\left|  {\rm val} \right\rangle \\
\frac{1}{N_{c}}\mathcal{J}_0\left(  \bm{z}\right)   & =&\sum_{m^{0}}\frac
{1}{\varepsilon_{{\rm val} }-\varepsilon_{m^{0}}}\left\langle {\rm val} \right.  \left|
\bm{z}\right\rangle \bm{\sigma}\cdot\bm{\tau}\left\langle
\bm{z}\right.  \left|  m^{0}\right\rangle \left\langle m^{0}\right|
\gamma^{0}\left|  {\rm val} \right\rangle
\nonumber\\ &&
+\sum_{n,m^{0}}\mathcal{R}_{4}(\varepsilon_{n},\varepsilon_{m^{0}
})\left\langle n\right.  \left|  \bm{z}\right\rangle \bm{\sigma}
\cdot\bm{\tau}\left\langle \bm{z}\right.  \left|  m^{0}\right\rangle
\left\langle m^{0}\right|  \gamma^{0}\left|  n\right\rangle
\end{eqnarray}
with the regularization functions
\begin{eqnarray}
\mathcal{R}_{1}\left( \varepsilon _{n}\right) &=& -\frac{\varepsilon
  _{n}}{2 \sqrt{\pi }}\int_{1/\Lambda ^{2}}^{\infty }\frac{du}{\sqrt{u}}
e^{-u\varepsilon _{n}^{2}}, \cr
\mathcal{R}_{2}\left( \varepsilon _{n},\varepsilon
_{m}\right)&=&\frac{1}{2}\frac{\mathrm{sgn}(\varepsilon _{m})
-\mathrm{sgn}(\varepsilon _{n})}{\varepsilon _{n}-\varepsilon
_{m}},  \cr
 \mathcal{R}_{3}(\varepsilon _{n},\varepsilon _{m})
&=&\frac{1}{2\pi }\int_{1/\Lambda ^{2}}^{\infty }dw\,\int_{0}^{1}d\beta \,%
\frac{\left( 1-\beta \right) \varepsilon _{n}-\beta \varepsilon _{m}}{\sqrt{%
\left( 1-\beta \right) \beta }}e^{-\left[ \varepsilon _{n}^{2}\left( 1-\beta
\right) +\varepsilon _{m}^{2}\beta \right] w},\cr
\mathcal{R}_{4}(\varepsilon _{n},\varepsilon _{m}) &=&
\frac{1}{2\sqrt{\pi }}\int_{1/\Lambda ^{2}}^{\infty }
\frac{du}{\sqrt{u}}\frac{\varepsilon _{m}e^{-\varepsilon
_{m}^{2}u}-\varepsilon _{n}e^{-\varepsilon _{n}^{2}u}}{\varepsilon
_{n}-\varepsilon _{m}} .
\end{eqnarray}

\end{appendix}

\end{document}